\documentclass[twocolumn]{aastex631}

\usepackage{amsmath,amstext}
\usepackage{stackengine}
 \usepackage{mathtools}

\usepackage{graphicx}
\usepackage{txfonts}
\usepackage{xcolor}
\usepackage{comment}

\newcommand{\noop}[1]{}

\newcommand{\ME}{\mbox{$M_{\Earth}$}} 

\newcommand{\be}{\begin{eqnarray}}
\newcommand{\ee}{\end{eqnarray}}

\newcommand{\MSun} {\mbox{$M_{\odot}$}}

\graphicspath{{./}{figures/}}

\shorttitle{Stellar flyby -- outer solar system}
\shortauthors{Pfalzner, Govind \& Wagner}

\begin{document}

\title{Irregular moons possibly injected from the outer solar system by a stellar flyby}

\author[0000-0002-5003-4714]{Susanne Pfalzner} 
\affiliation{J\"ulich Supercomputing Center, Forschungszentrum J\"ulich, 52428 J\"ulich, Germany}
\affiliation{Max-Planck-Institut f\"ur Radioastronomie, Auf dem H\"ugel 69, 53121 Bonn, Germany}

\author[0000-0002-3663-1168]{Amith Govind}
\affiliation{J\"ulich Supercomputing Center, Forschungszentrum J\"ulich, 52428 J\"ulich, Germany}
\affiliation{Department of Physics, University of Cologne, 50923 Cologne, Germany}

\author[0009-0001-5904-1742]{Frank W. Wagner}
\affiliation{J\"ulich Supercomputing Center, Forschungszentrum J\"ulich, 52428 J\"ulich, Germany}

\begin{abstract}
The irregular moons orbit the giant planets on distant, inclined, and eccentric trajectories, in sharp contrast with the coplanar and quasicircular orbits of the regular moons. The origin of these irregular moons is still an open question, but these moons have a lot in common with the objects beyond Neptune (trans-Neptunian objects -- TNOs), suggestive of a common origin. Here, we show that the close flyby of a star may be the connecting element. A stellar flyby can simultaneously reproduce the complex TNO dynamics quantitatively while explaining the origin of the irregular moons and the colour distributions of both populations. This flyby would have catapulted 7.2~\% of the original TNO population into the region of the planets, many on retrograde orbits. Most injected TNOs would have been subsequently ejected from the solar system (85~\%). However, a considerable fraction would have had the potential to be captured by the planets. 
The exclusively distant origin of the injected TNOs may also explain the lack of very red irregular moons.
\end{abstract}


\section{Introduction} 
\label{sec:intro}

The discovery of irregular moons around the outer planets (Jupiter, Saturn, Uranus, and Neptune) may provide invaluable insights for understanding the formation and evolution of the solar system. These moons exhibit unique features, such as orbiting their host planets on more distant, inclined, and eccentric orbits than the regular moons. Interestingly, the irregular moons may have been captured from the region beyond Neptune or from accretion regions near the giant planets shortly after planet formation \citep{Jewitt:2007}. Therefore, studying irregular moons could help us verify or rule out different solar system formation and evolution hypotheses.

The question is how the TNOs moved from their original location beyond Neptune into the giant planet region. One hypothesis suggests that the irregular moons were injected when the outer planets migrated in the planetesimal disc \citep{Nesvorny:2007,Nesvorny:2014}. 

Here, we present an alternative hypothesis where a stellar flyby could have caused objects outside Neptune's orbit to enter the planet region. Previous work suggested a flyby as an add-on to planet scattering to explain certain dynamic families \citep{Kobayashi:2005,Kenyon:2004,Morbi:2005}. By contrast, in \citet{Pfalzner:2024} we showed that a flyby can explain on its own the diversity of the TNOs' dynamics. We test the same flyby that closely reproduces the different TNO dynamic groups on a quantitative level for its ability to inject TNOs into regions where the giant planets may have captured them. We will see that this model's strength is its simplicity.  It requires only one process to simultaneously convincingly explain the irregular moon's features and the TNO dynamics.

\begin{table*}[ht]
\centering
\caption{Irregular moon populations of the giant planets.}
\label{tab:moons}
\begin{tabular}{l|cc|cll|cll|l}
\hline\hline
\multicolumn{3}{c}{} & \multicolumn{3}{c}{prograde} &  \multicolumn{3}{c}{retrograde} &  \\
planet & No. & No. groups (sum/p/r) & No. & min $i$ (°) & max $i$ (°)& No. & min $i$ (°) & max $i$ (°) & source \\
\hline
Jupiter &  87   & 5+2/4/3  &  16 & 26.63 & 53.2 &  71 & 166.7 & 139.8 &  \citet{Sheppard:2023}\\
Saturn  &  122  & ?? &  22  & 33.82 & 52.46 & 100 & 179.8 & 145.8 & \citet{Sheppard:2023}\\
Uranus  &  10  & ??  & 1 & 56.6 & 56.6 & 9 & 169.8 & 140.9 & \citet{Irregular_moons}\\
Neptune &  9  & ?? &  4 & 7.23 & 48.51 & 5 & 156.8 & 132.6 & \citet{Irregular_moons}\\
\hline
\end{tabular}
\end{table*}

\section{Observational constraints}

According to observations (see Tab.~\ref{tab:moons}), Jupiter has 87 irregular moons, while Saturn has 122. Therefore, it seems that Saturn has more irregular moons than Jupiter. Additionally, Uranus has ten irregular moons, and Neptune has nine irregular moons. It is important to note that these numbers are likely to be higher due to the increasing difficulty of detecting moons at larger distances.

Interestingly, most irregular moons orbit on retrograde orbits. The retrograde to prograde ratios are 71:16 for Jupiter, 100:22 for Saturn, 9:1 for Uranus, and 5:4 for Neptune \citep{Sheppard:2023, Brozovic:2022}. However, the irregular moon populations likely have evolved since their formation, with the moons believed to be fragments of once larger parent bodies \citep{Bottke:2010}. Therefore, comparing the number of prograde and retrograde families may be a more accurate approach. For Jupiter, three retrograde and two prograde groups have been identified, indicating that the dominance of the retrograde moons is significantly less pronounced for the families (3:2) than that of Jupiter's individual moons (71:16). For Saturn, the number of moon families has not been determined yet because its retrograde moons are not well-clustered. For Uranus and Neptune, the low number of known objects does not allow for determining family ratios.

While the irregular moons have characteristic high inclinations, there is a lack of objects with inclinations in the range 55~° -- 130~° \citep{Carruba:2002}. The reason is unclear. It could be due to primordial factors, the capture processes, or long-term evolution. Planetary and solar perturbations could have affected the moons' orbits, increasing their eccentricities while decreasing their inclinations through Kozai resonance \citep{Carruba:2002, Nesvorny:2002}.

The irregular moons exhibit different colours, ranging from grey to red.  For a subset of these moons, \citet{Belyakov:2024} find that their spectrophotometry matches spectra of some Neptune trojans and excited Kuiper belt objects, suggesting shared properties. However, the irregular moons tend to be deficient in very red objects compared to the TNOs \citep{Jewitt:2007,Pena:2022}. The cause of this observation is not yet fully understood.

\section{Method}
\label{sec:method}

We model the flyby of a 0.8~\MSun\ star at a periastron distance to the Sun of \mbox{$r_p=$ 110 au} at an inclination of \mbox{$i_p=$ 70 °}. Based on a previous study, this stellar flyby appears to be the best match for the orbital parameters of the currently known TNOs, including the retrograde and Sedna-like objects \citep{Pfalzner:2024}.
Here, we focus on the test particles representing TNOs pushed into the region inside Neptune's orbit (\mbox{$r_p^f\lessapprox$ 29 au}). At this stage, the gas masses in circumstellar discs are small (\mbox{$m_d\ll$ 0.1 \MSun}), so that viscosity can be neglected. Therefore, the problem reduces to gravitational three-body interactions between the Sun, the perturber star, and each test particle \citep[e.g.,][]{Kobayashi:2001, Musielak:2014, Pfalzner:2018}.

We use the integrator IAS15 of the REBOUND code \citep{Rein:2014} to compute with great precision the effect of the stellar flyby on the planetesimal disc surrounding the Sun. This precision is required for the consecutive investigation of the long-term behaviour of the injected particles, which we perform in a second step.

The primordial disc is modelled by 10$^5$ massless test particles initially orbiting the Sun on circular orbits. We consider two disc sizes, 150~au and 300~au, as sizes of 100~au to 500~au have been observed to be typical for protoplanetary and debris discs \citep{Andrews:2020,Hendler:2020}. We employ a constant test particle surface density to obtain a high resolution in the outer disc regions. Individual masses are assigned to the particles to model the associated mass density distribution afterwards \citep{Hall:1997, Steinhausen:2012}. We did not include the giant planets in the simulation of the flyby itself, but in the modelling of the long-term evolution.

The simulation of the long-term evolution start when stable orbits are established about 12~kyr after periastron passage. 
Using the GENGA code \citep{Grimm:2014, Grimm:2022} along with its hybrid symplectic integrator \citep{Chambers:1999}, we study the long-term trajectories of injected TNOs from the positions and velocities of the test particles at $t$ = 12 kyr.  Although GENGA is optimised for studying long-term evolution, these simulations are computationally expensive. Therefore, we limited our simulations to the first 1~Gyr after the flyby. By gradually reducing the time step, we determined that a time step of \mbox{$\Delta t=$ 30 d} is a good choice to efficiently obtain the required temporal resolution. 

\begin{figure}
\centering
\includegraphics[width=0.45\textwidth]{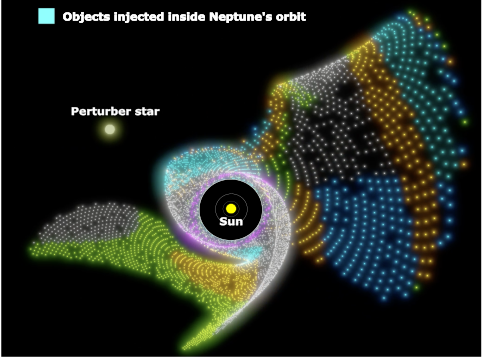}
\caption{Snapshot of stellar flyby. Test particle locations at \mbox{$t=$ 200 yr} after periastron passage of perturber star. The turquoise particles indicate the TNOs injected into the planet region by the flyby. The perturber star passed through the disc at a perihelion distance of 110~au on the right-hand side of the picture.}
\label{fig:flyby}
\end{figure}

\section{Results}

During a flyby, the particles in front of the perturber star slow down, while the ones behind speed up, creating characteristic spiral arms \citep[see,][]{Pfalzner:2003}. If the star passes on an inclined orbit, like in our case,  the structure becomes more complex, as shown in Fig.~\ref{fig:flyby}. There, the test particles that will end up on orbits with periastron distances \mbox{$r_p^f<$ 29 au} are indicated in turquoise. These particles have sub-Keplerian velocities and are excited onto highly eccentric orbits with small perihelion distances. Initially, all injected TNOs were at large distances from the Sun (\mbox{$r_p^i>$ 60 au}), primarily on the side of the disc where the perturbing star passed.

\begin{figure}
\centering
\includegraphics[width=0.45\textwidth]{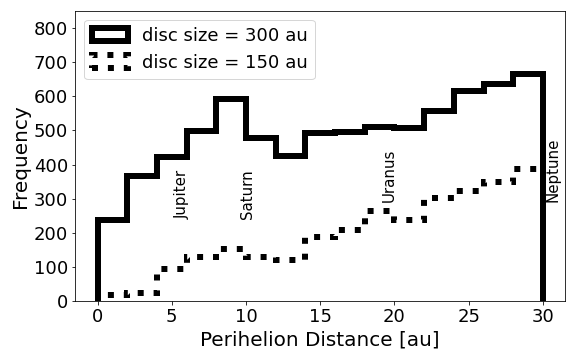}
\includegraphics[width=0.45\textwidth]{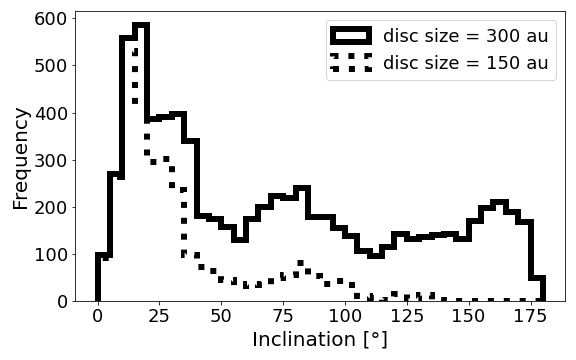}
\caption{Distributions of the injected TNOs as a function of periastron distance (top) and inclination (bottom). The solid line shows the results for a primordial disc extending to 300~au and the dashed line for a 150~au primordial disc.}
\label{fig:disc_sizes}
\end{figure}

Around 7.2~\% of the original disc, corresponding to 12.9~\% of the final bound population, were injected inside 29~au. The total mass of injected TNOs lies in the range \mbox{0.015 -- 1.8~\ME,} roughly corresponding to 1 -- 146 times the mass of the Moon. This corresponds to 90 -- 10~000 objects larger than 1000~km, to 90~000 -- 10~000~000 objects of a size of 100~km, or to many more smaller objects. These estimates are based on the minimum mass of debris discs \citep[\mbox{$M_d=$ 5 -- 20 \ME,}][]{Krivov:2021} and the typical density of objects like Pluto. For comparison, the total mass of the known TNO population is estimated to be $\approx$0.02~\ME\ \citep{Pitjeva:2018}.

Figure~\ref{fig:disc_sizes} depicts the periastron distance $r_p^i$ and inclination $i_p^i$ distributions of the injected TNOs after the flyby for primordial disc sizes of 150 and 300 au, respectively. TNOs are injected into the entire planet region (see Fig.~\ref{fig:disc_sizes}, top). Smaller disc sizes (dashed distribution) clearly lead to fewer TNOs being injected, and the region inside Jupiter is less enriched in injected TNOs. Nonetheless, even for a 150~au-sized disc, several TNOs end up on orbits with perihelia of down to 0.2~au, which is well inside Mercury's orbit. For the 300~au-sized disc, periastron distances as small as 0.07~au can be found.

Saturn's larger number of irregular moons than Jupiter's could be due to more TNOs being injected in this region by the flyby (see Fig.~\ref{fig:disc_sizes}, top). Our simulations show a similar number of injected TNOs into the regions of Uranus and Saturn. However, Uranus's cross-section for capture is smaller. Hence, we expect the total matter captured by Uranus to be approximately a third of Saturn's. Currently, the detection sensitivity for the outermost planets is low, which hinders a meaningful test of this prediction. However, it may be possible to test this hypothesis when LSST becomes operational \citep{Ivezic:2019}.





\begin{figure}
\centering
\includegraphics[width=0.45\textwidth]{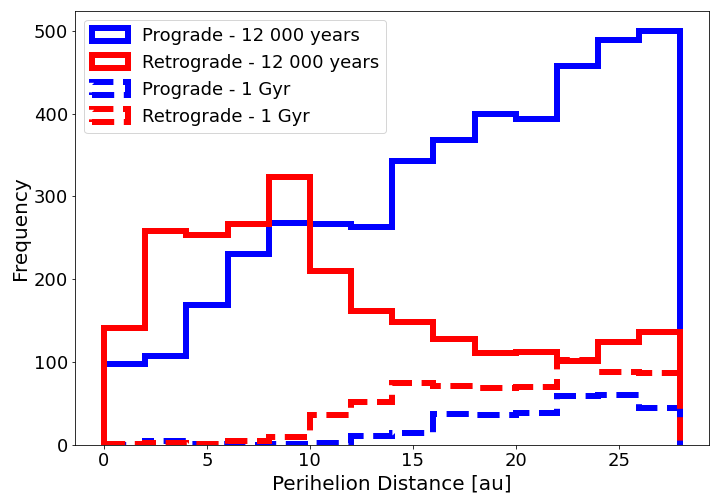}
\includegraphics[width=0.45\textwidth]{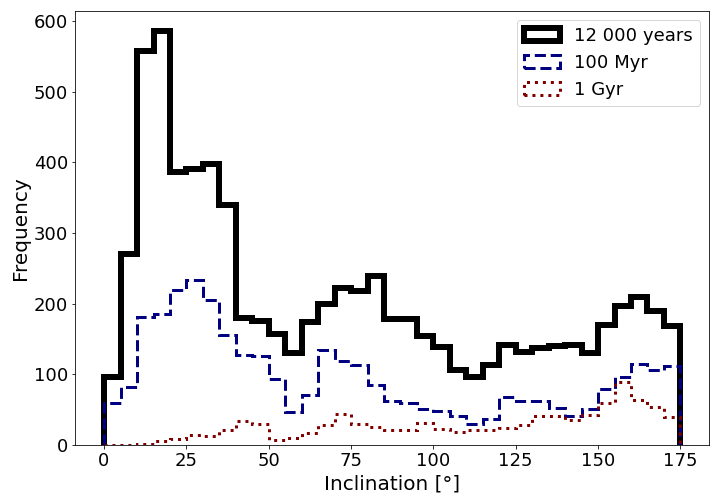}
\caption{Evolution of the injected TNO population over the first 1~Gyr after the stellar flyby (300~au initial disc size). Top: Distribution of prograde (blue lines) and retrograde (red lines) TNOs as a function of periastron distance at 12~kyr (solid lines) and 1~Gyr (dashed lines). Bottom: Distribution of inclination for the test particles remaining bound at 12~kyr, 100~Myr, and 1~Gyr, respectively.}
\label{fig:evolution}
\end{figure}

Surprisingly, a significant number of TNOs are injected on high inclination or even retrograde orbits (see Fig.~\ref{fig:disc_sizes}, bottom). While most TNOs are injected on prograde orbits inclined by 10~° to 20~°, retrograde-injected TNOs outnumber the prograde ones in the region inside approximately 10 au (see Fig.~\ref{fig:evolution}, top, solid lines). In the Jupiter region, TNOs are about 30~\% more likely to be injected on a retrograde than a prograde orbit. Near Saturn, retrograde injected TNOs are 20~\% more likely than prograde ones. The flyby scenario is the first model that would explain the dominance of retrograde objects due to the injection mechanism, at least for the regions around Jupiter and Saturn.

Nevertheless, in our simulation, the dominance of the retrogrades is less pronounced compared to the observed individual moons. The reason likely is that the detected moons are fragments of once larger bodies. If one compares the ratio for the dynamical families of Jupiter, there seems to be good agreement. Unfortunately, similar constraints are missing for the other outer planets. 


\begin{figure}
\centering
\includegraphics[width=0.45\textwidth]{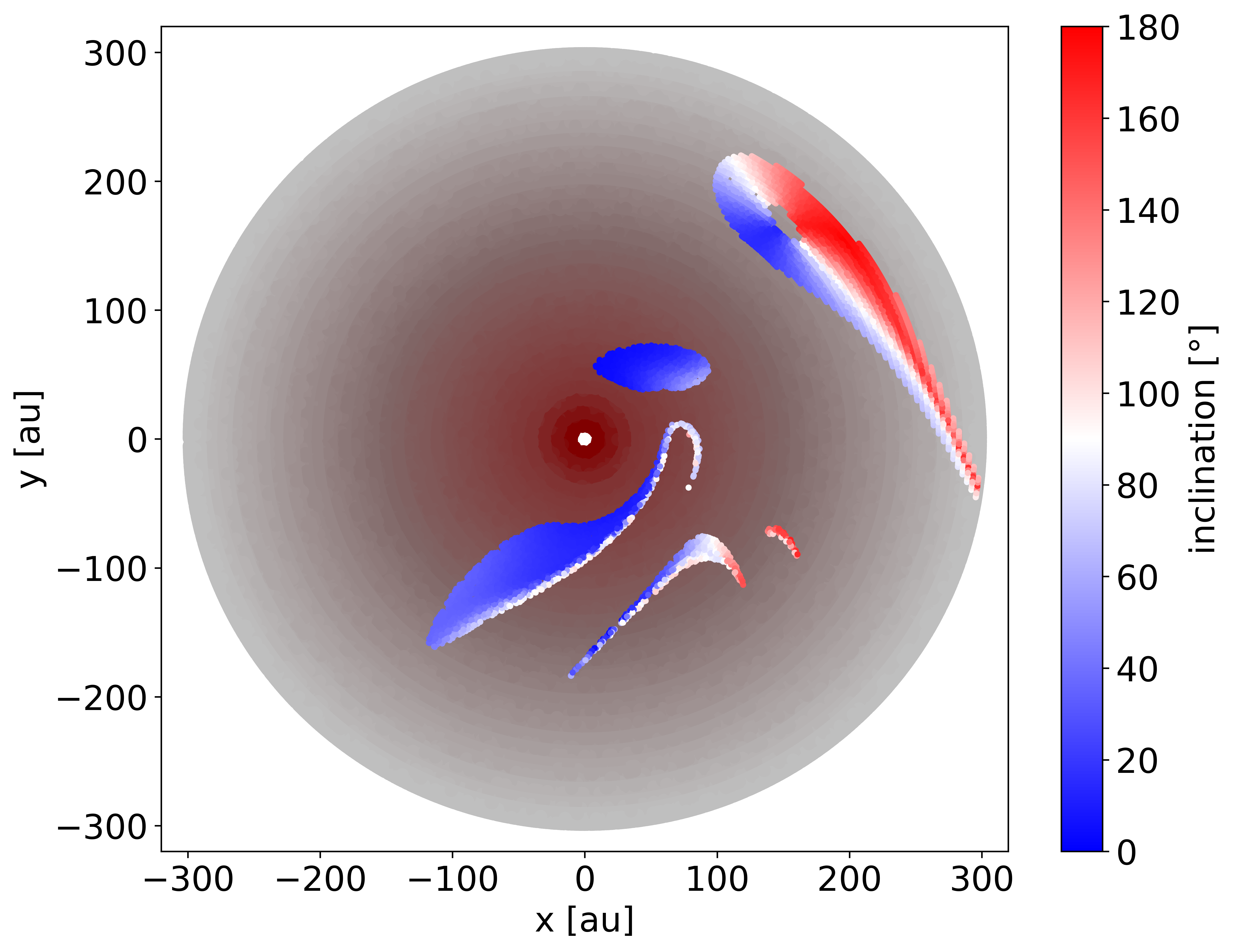}
\includegraphics[width=0.45\textwidth]{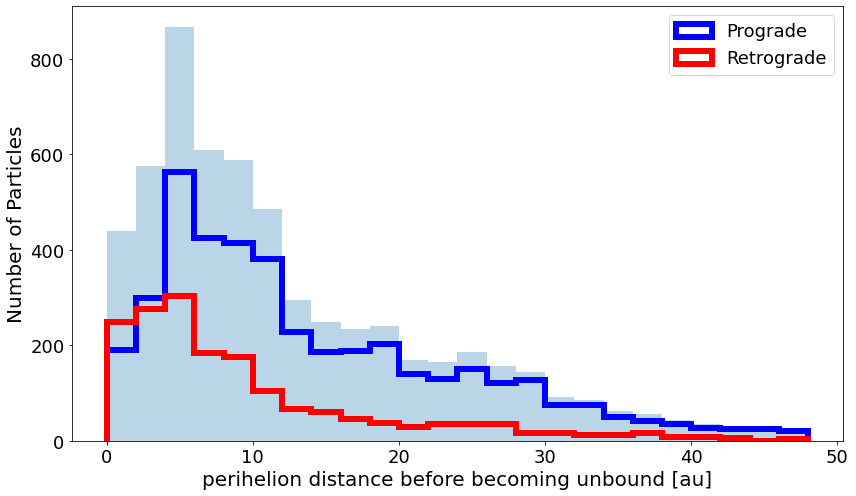}
\caption{Top: Origin of the injected TNOs in the unperturbed disc of 300~au in size. The disc has a colour gradient from reddish to greyish corresponding to the colour of planetesimals. The patches represent the feeding zones or origin of the injected TNOs. The colourbar gives the inclination of the injected TNOs; blue represents prograde objects and red represents retrograde objects. Bottom: Distribution of periastron distance of the ejected particles just before becoming unbound for the prograde (blue), retrograde (red), and total population (grey).}
\label{fig:origin}
\end{figure}

In Fig.~\ref{fig:origin}, top, we can see the origin of the injected TNOs overlaid onto a pre-flyby disc with a colour gradient ranging from very red to grey. This choice is based on the cold Kuiper belt objects being predominantly very red in colour and likely residing close to their original position in the disc {\bf \citep{Lacerda:2014}}. Since the injected TNOs all originate from the region beyond 60~au, we would expect TNO colour to cover the entire spectrum from red to grey, but, importantly, lacking very red TNOs. Thus, the observed lack of very red objects among the irregular moons would be a direct consequence of the origin of the injected TNOs in the outer regions of the disc. Currently, we are investigating the TNO colours. Our preliminary results suggest that a flyby may also be able to explain the observed correlation between TNO colours and inclination \citep{Schwamb:2019}.



A flyby provides explanations for many of the irregular moons' features. However, there is one exception. While our simulations provide an almost uniform distribution of TNOs' inclination between 50~° and 180~° (see Fig.~\ref{fig:disc_sizes}, bottom), observations show a lack of moons with inclinations in the range 55~° -- 130~°. Likely, the absence of close-to-polar orbiting moons is a result of the long-term evolution of the orbits after capture \citep{Carruba:2002, Nesvorny:2002}.
The Kozai mechanism sets an upper limit of $\approx$60~° on their inclinations. Any moon whose inclination exceeded this value would periodically gain an eccentricity so large that its closest approach would take it into the region of the regular moons. Its fate would be a collision with one of them or with the host planet or an ejection from the host planet system \citep{Jewitt:2007}.


So far, we have looked at the situation immediately after the flyby, but the irregular moons might have been captured over a longer period. \cite{Punzo:2014} looked at the long-term evolution, but the study was limited to the first 100~Myr and the population outside Neptune's orbit. Little is known about the long-term fate of the injected TNOs. Here, we study the dynamics over the first Gyr (see Fig.~\ref{fig:evolution}). Over this period, 85~\% of the TNOs injected into the planet region are ejected from the solar system. Jupiter is the most efficient ejector, expelling nearly 98~\% of the TNOs present in the 4~au to 6~au region, compared to only 77~\% at the location of Uranus (18~au -- 20~au) (see Fig.~\ref{fig:origin}, bottom).

However, retrograde-injected TNOs are less often ejected than prograde ones (see Fig.~\ref{fig:evolution}, top).  Almost 24~\% of the original total retrograde population is retained. Even retrogrades near the plane (near 180~°) have a reasonable chance of a long-term presence in the planet region (see Fig.~\ref{fig:evolution}, bottom).



\section{Discussion}

One question is whether the giant planets could have captured the injected TNOs. Our flyby simulations cannot directly model the capturing of injected TNOs by the planets because we do not model individual TNOs but test particles representative of an entire group of TNOs.  Besides,  a much higher resolution of the disc would be required for a sufficient number of test particles close to the planets to resolve the surroundings of the planets. Additionally, the modelling of the evolution of these high-resolution simulations over Gyr timescales is prohibitively computationally expensive. 

Nonetheless, there is good reason to believe these injected TNOs could have been captured. In their research on the planet instability (`Nice Model'), \cite{Nesvorny:2007,Nesvorny:2014} explored the possibility that the irregular moons of the giant planets were captured from a circumstellar planetesimal disc by three-body gravitational interactions.  By injecting numerous objects from the TNO region, \cite{Nesvorny:2007} found that planetary encounters can create moons on distant orbits at Saturn, Uranus, and Neptune. This capture process is efficient enough to produce populations of irregular moons. The trajectories of the injected objects in this planet scattering approach and those injected by a flyby are similar. Therefore, the TNOs injected by flybys should have a similar chance of being captured.

The original model by \cite{Nesvorny:2007} had the problem that it could not explain the irregular moons around Jupiter, as Jupiter itself did not often take part in planetary encounters. In later versions (`Jumping Jupiter Model'), \cite{Nesvorny:2014} assumed that the early solar system contained at least one additional giant planet. This allowed the planet scattering process to account for Jupiter's irregular moons. However, this improved version still has difficulties in explaining the observed asymmetry between prograde and retrograde irregular moons and the lack of very red objects among them. The flyby scenario overcomes this problem.

\citet{Jewitt:2007} suggested that the asymmetry between prograde and retrograde irregular moons could be at least partly due to long-term effects. Retrograde moons are more stable than their prograde counterparts, especially on highly elongated orbits \citep{Hamilton:1991,Hamilton:1997,morais:2012}. Another point to consider is that detecting prograde moons is more challenging than observing retrograde moons \citep{Ashton:2021}, leading to a bias towards detecting retrograde moons. Currently, the relative importance of these effects is not well-constrained, and their influence on the abundance of retrograde objects is an open question.

\section{Summary and Conclusions}

This study explores the possibility of irregular moons being TNOs that were injected into regions close to the giant planets by a nearby stellar flyby and subsequently captured by the planets. The strength of this model is that it can explain features of both regions beyond \citep{Pfalzner:2024} and within Neptune's orbit. We start with the specific flyby that can reproduce the diversity of the orbits of objects beyond Neptune, including high-inclination, Sedna-like, and retrograde bodies. We find that such a flyby would have catapulted 7.2~\% of the original TNO population into regions close to the giant planets, many on retrograde orbits. A large portion of these objects would have been subsequently ejected from the solar system, but a considerable fraction would have remained for extended periods in the catchment area of the planets.

The stellar flyby could account for several unexplained features. It would have led automatically to more moons around Saturn than Jupiter, and retrograde moons would dominate around both planets. The exclusive origin of injected TNO beyond 60~au would also explain the lack of very red irregular moons. However, one puzzle remains regarding the absence of irregular moons with close-to-polar orbits, which might be related to the capture process or long-term effects.

Our results imply that it may be worthwhile to further investigate the possible common origin of irregular moons and trojans as suggested by \cite{Nesvorny:2002,Graykowski:2018} from a flyby perspective. The injected TNOs may have also transported volatiles and prebiotics to the rocky inner planets \citep{Chyba:1990,Todd:2020,Anslow:2023}. Thus, while the flyby would have left the planets' orbits completely untouched, it may have played a role in the emergence of life. However, how much prebiotic material originally contained in an injected TNO would survive impact on a terrestrial planet would require further studies.

\begin{acknowledgments}We thank the referee for the valuable suggestions on improving the manuscript. We are also grateful to Nadar Haghighipour for pointing out the relevance of the Kozai mechanism on the inclination of the irregular moons. SP acknowledges funding for this project through grant no. 450107816 of the Deutsche Forschungsgemeinschaft.\end{acknowledgments}

\bibliography{reference}{}
\bibliographystyle{aasjournal}

\end{document}